# Structure-Composition-Property Relationships in Antiperovskite Nitrides: Guiding a Rational Alloy Design

Hongxia Zhong[1,2], Chunbao Feng[3], Hai Wang[1], Dan Han[4], Guodong Yu[2], Wenqi Xiong[2], Yunhai Li[2], Mao Yang[5,6], Gang Tang[7*], and Shengjun Yuan[2*]

[1]School of Mathematics and Physics, China University of Geosciences (Wuhan), Wuhan 430074, China

[2]School of Physics and Technology, Wuhan University, Wuhan, 430072, People's Republic of China

[3]School of Science, Chongqing University of Posts and Telecommunications, Chongqing 400065, China

[4]Department of Chemie, Ludwig-Maximilians-Universität München, München, 81377, Germany

[5]Institut für Physik and IRIS Adlershof, Humboldt-Universität zu Berlin, Berlin 12489, Germany

[6]School of Science, Xi'an Polytechnic University, Xi'an 710048, P. R. China

[7]Theoretical Materials Physics, Q-MAT, CESAM, University of Liège, B-4000 Liège, Belgium

**Corresponding authors**
*E-mail: gtang@uliege.be and sjyuan@whu.edu.cn

**Abstract**

The alloy strategy through A- or X-site is a common method for experimental preparation of high-performance and stable lead-based perovskite solar cells. As one of the important candidates for lead-free and stable photovoltaic absorber, the inorganic antiperovskite family has recently been reported to exhibit excellent optoelectronic properties. However, the current reports on the design of antiperovskite alloys are rare. In this work, we investigated the previously overlooked electronic property (e.g., conduction band convergence), static dielectric constant, and exciton binding energy in inorganic antiperovskite nitrides by first-principles calculations. Then, we reveal a linear relationship between tolerance factor and various physical quantities. Guided by the established structure-composition-property relationship in six antiperovskite nitrides $X_3NA$ ($X^{2+} = Mg^{2+}, Ca^{2+}, Sr^{2+}$; $A^{3-} = P^{3-}, As^{3-}, Sb^{3-}, Bi^{3-}$), for the first time, we design a promising antiperovskite alloy $Mg_3NAs_{0.5}Bi_{0.5}$ with the quasi-direct band gap of 1.402 eV. Finally, we make a comprehensive comparison between antiperovskite nitrides and conventional halide perovskites for pointing out the future direction for device applications.

# Table of Contents

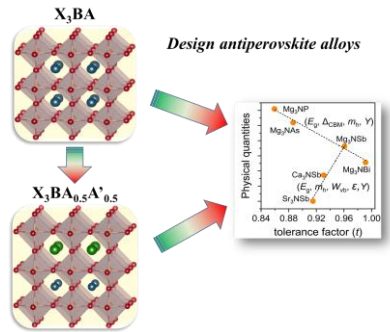

**Introduction**

Perovskite semiconductors have attracted intensive research interest in recent years due to their versatile compositions and tunable electronic and optical properties.[1-3] As an important member in the perovskite family, lead (Pb) halide perovskites APbX$_3$ (A$^+$ = CH$_3$NH$_3^+$, CH(NH$_2$)$_2^+$, Cs$^+$; X$^-$ = I$^-$, Br$^-$, Cl$^-$) have achieved great success of applications in solar cells,[1] light-emitting diodes,[2] and photodetectors[3, 4] because of their intriguing optoelectronic properties (see Figure 1). For example, the power conversion efficiency (PCE) of perovskite solar cells has been rapidly improved from 3.8% to 25.5% within only one decade.[5, 6] Despite the exciting progress, long-term instability and Pb toxicity are still key challenges hindering perovskite solar cells from practical applications.[7]

An effective strategy to solve the above problem is to carry out cation or anion substitution. For example, the most straightforward way to eliminate the toxic Pb(II) is to replace it with Sn(II). The obtained ASnX$_3$ has a crystal structure and electronic properties similar to Pb counterparts, but unfortunately they suffer from the serious instability issues (i.e., against the oxidation to Sn$^{4+}$).[8] When replacing I$^-$ with S$^{2-}$/Se$^{2-}$, it can lead to chalcogenide perovskites ABX$_3$ (A$^{2+}$ = Ca$^{2+}$, Sr$^{2+}$, Ba$^{2+}$; B$^{4+}$ = Ti$^{4+}$, Zr$^{4+}$, Hf$^{4+}$; X$^{2-}$ = S$^{2-}$, Se$^{2-}$) with robust thermal stability and nontoxic elements (see Figure 1).[9, 10] Due to the large electronegativity difference between B- and X-site ions and the disordered ground-state structure, they usually exhibit a large band gap, and they are difficult to form high-quality films at low temperatures.[9] Although the experimentally synthesized Ba$_3$Zr$_2$S$_7$ posses a surprisingly optimal band gap of 1.28 eV, its valence band maximum (VBM) and conduction band minimum (CBM) do not exhibit sufficient dispersion (i.e., small band width $W$ and large carrier effective mass $m$).[11] Therefore, it is very challenging to realize the well-performed solar cell based on chalcogenide perovskites.

Another effective strategy is to reverse the ion type of perovskite lattice sites. Specifically, through electronically inverting the formula ABX$_3$ (A and B are cations, and X is an anion), a class of antiperovskite X$_3$BA can be obtained, in which anions occupy the A and B sites, and cations occupy the X sites (see Figure 1).[12] Recently, Gebhardt *et al.* proposed a series of inverse-hybrid perovskites (CH$_3$NH$_3$)$_3$BA (B: monovalent anions, A: divalent anions or B: divalent anions, A: monovalent anions) for photovoltaic applications. Unfortunately, these predicted compounds still do not exhibit the dispersive valence and conduction bands near the Fermi level.[13, 14] At present, several synthesized hybrid organic-inorganic antiperovskites have shown potential application in the field of molecular ferroelectric,[15] but no hybrid antiperovskite suitable as a photovoltaic absorber has been experimentally reported. On the contrary, all-inorganic antiperovskite semiconductors (i.e., X$_3$NA, X$^{2+}$ = Mg$^{2+}$, Ca$^{2+}$, Sr$^{2+}$, Ba$^{2+}$; A$^{3-}$ = P$^{3-}$, As$^{3-}$, Sb$^{3-}$, Bi$^{3-}$) have been reported to exhibit attractive electronic structure, optical absorption, ion migration barriers, and defect properties, which show great potential applications in photovoltaics, thermoelectrics, and solid electrolytes (see Figure 1).[16-18] For example, Dai[17] and Mochizuki[18] have respectively proposed Ca$_3$NSb, Ca$_3$NBi, Mg$_3$NP, and Sr$_3$NP as promising photovoltaic absorbers because of their suitable direct band gaps and high optical absorption coefficients. Although the antiperovskite family greatly expands the composition space for searching Pb-free and stable photovoltaic candidates, there are few reports that establish their structure-composition-property relationships in detail. Moreover, as far as we know, no specific strategy has been reported for the design of antiperovskite alloys for photovoltaic applications.

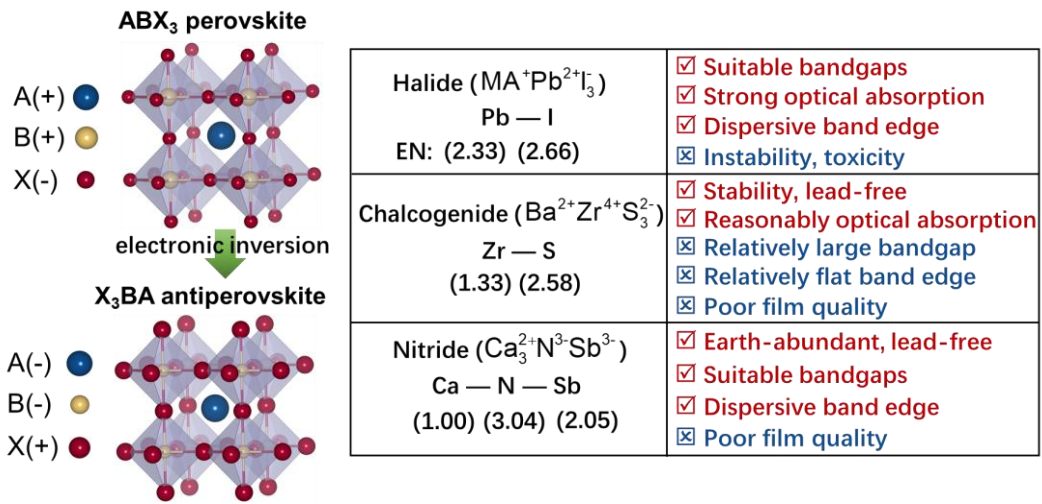

Figure 1: Crystal structures of cubic perovskite ABX$_3$ and antiperovskite X$_3$BA. Three representatives, lead halide perovskite, chalcogenide perovskite, and antiperovskite nitride, are presented, with their corresponding preponderances and disadvantages. The electronegativities

(EN) of the elements are given in parentheses.

In this work, we first investigated the electronic property, static dielectric constant, exciton binding energy of six inorganic antiperovskite $X_3NA$ ($X^{2+}$ = $Mg^{2+}$, $Ca^{2+}$, $Sr^{2+}$; $A^{3-}$ = $P^{3-}$, $As^{3-}$, $Sb^{3-}$, $Bi^{3-}$) to establish the structure-composition-property relationships. Based on the calculated results, a general linear relationship between the tolerance factor and various physical quantities is uncovered. Guided by the structure-composition-property relationship, we design a promising antiperovskite alloy $Mg_3NAs_{0.5}Bi_{0.5}$ with a quasi-direct band gap of 1.402 eV. Finally, we make a comprehensive comparison between antiperovskite nitrides and conventional lead halide perovskites. Our studies will provide a strategy to design favorable antiperovskite alloys for novel device applications.

**Computational Details**

Our calculations are performed using the projector augmented wave (PAW) method implemented in the Vienna ab initio simulation package (VASP) code.[19, 20] The standard PAW pseudo potentials are adopted.[21] We use the Perdew, Burke, and Ernzerhof (PBE) form of the generalized gradient approximation (GGA) exchange correlation functional[22] for crystal structure relaxation, and the hybrid density functional (Heyd-Scuseria-Ernzerhof, HSE06)[23] for electronic properties of antiperovskites. The cut-off energy is set to 500 eV after convergence tests. We employ 3×3×3, 6×6×6, and 9×9×9 Γ-centered Monkhorst-Pack $k$-point[24] grid for relaxations, self-consistent calculations, and density of states (DOSs) calculations, respectively. In our current calculations, the total energy is converged to less than $10^{-5}$ eV. The maximum force is less than 0.02 eV/Å during the optimization, where both the lattice constants and atomic positions are fully relaxed. The spin–orbit coupling (SOC) effects have been included for the electronic band structures of all studied antiperovskites $X_3BA$. The phonon dispersion calculations are based on PHONOPY code.[25] A 8×3×1 supercell with 13×9×1 k mesh is used to ensure the convergence.

For cubic antiperovskites, there are three nonzero elastic stiffness constants $C_{11}$, $C_{12}$ and $C_{44}$, and the stress-strain relationship is obtained from Hooke's law under plane-stress condition $E_i = C_{ij} \varepsilon_i$. Here, we recalculate the strained sample using higher cut-off energy, and get the elastic constants $C_{ij}$ using the VASPKIT code.[26] Then, the Bulk modulus $B$, shear modulus $G$, Young's modulus $Y$, and Poisson's ratio $\upsilon$ can be derived as[27]

$$B = \frac{C_{11} + 2C_{12}}{3}$$
$$G = \frac{C_{11} - C_{12} + 3C_{44}}{5}$$
$$Y = \frac{9BG}{3B + G}$$
$$\upsilon = \frac{3B - 2G}{2(3B + G)}$$

The dielectric properties of the studied compounds are calculated by density functional perturbation theory (DFPT).[28, 29] In DFPT, the dielectric constant tensor is defined as a linear response to the perturbative electric field and the ionic displacement is considered as a perturbation to the equilibrium system. Subsequently, the static dielectric constant ($\varepsilon_{std}$) consists of the electronic ($\varepsilon_{ele}$) and the ionic parts ($\varepsilon_{ion}$) of the system to the applied electric field.[30] The $\varepsilon_{ele}$ is the response of the electronic charge density to the perturbative electric field in the linear response regime. In order to increase accuracy for the dielectric constant calculations, the total energy is converged to less than $10^{-6}$ eV, and the maximum force is less than 0.001 eV/Å.

Based on the calculated $\varepsilon_{ele}$, we estimate the exciton binding energy $E_b$ using the Wannier model[31]

$$E_b = R_y \frac{\mu^*}{\varepsilon^2}$$

where Ry = 13.56 eV is the atomic Rydberg energy, $\mu^*$ is the reduced exciton mass ($1/\mu^* = 1/m_e + 1/m_h$), and $\varepsilon$ is the electronic dielectric constant.

**Results and Discussions**

**The effects of X- and A-site elements on electronic properties**

For conventional $ABX_3$ perovskites, it is well known that the B- and X-site anions have major impacts on the electronic structures, while A-site cation does not contribute the band edge because of its highly ionic nature.[32, 33] But the effect of composition on the electronic properties in antiperovskites has not been fully understood.[34] Here, we take the six antiperovskite nitrides $X_3NA$ ($X^{2+} = Mg^{2+}, Ca^{2+}, Sr^{2+}$; $A^{3-} = P^{3-}, As^{3-}, Sb^{3-}, Bi^{3-}$) as example to study their composition-property relationship. The stability, optimized structure parameters, and electronic properties are summarized in Table 1 and Supporting Information (see Figures S1-S6 and Table S2), which are consistent with previous reports. Here, we mainly focus on tuning the convergence of the lowest unoccupied state between M and Γ point to realize direct-indirect/indirect-direct transition (see Figure 2a), which has been ignored in previous literatures. As shown in Figure 2a, we define quantitatively the energy difference between the two conduction states $Γ_c$ and $M_c$ point as $Δ_{CBM}$. The positive value of $Δ_{CBM}$ means the direct band gap, and the negative value means the indirect band gap. When changing the X-site element from Sr to Ca to Mg, the value of $Δ_{CBM}$ decreases from 1.858 eV in $Sr_3NSb$ to 1.696 eV in $Ca_3NSb$, to -0.214 eV in $Mg_3NSb$ (Table 2), meaning a direct-indirect transition. The change of $Δ_{CBM}$ also exists in $Mg_3NA$ ($A^{3-}$ = $P^{3-}, As^{3-}, Sb^{3-}, Bi^{3-}$) by substituting A-site elements. When replacing $Sb^{3-}$ with a heavier element (i.e., $Bi^{3-}$), $|Δ_{CBM}|$ becomes larger. However, when $Sb^{3-}$ is replaced with lighter elements (i.e., $As^{3-}$ and $P^{3-}$), the value of $Δ_{CBM}$ changes from negative to positive, suggesting that the direct band gap is restored.

Table 1: Calculated lattice constant $a$, Goldschmidt's tolerance factor $t$ (Table S1), band gap $E_g$, static dielectric constant $ε_{std}$, exciton binding energy $E_b$, and optical transition features between band edges at Γ point for $X_3NA$ ($X^{2+} = Mg^{2+}, Ca^{2+}, Sr^{2+}$; $A^{3-} = P^{3-}, As^{3-}, Sb^{3-}, Bi^{3-}$) antiperovskites.

| Compound | $a$ (Å) Calc. | $a$ (Å) Expt. | $t$ | $E_g$ (eV) | $ε_{std}$ | $E_b$ (meV) | Optical transition |
|---|---|---|---|---|---|---|---|
| $Mg_3NP$ | 4.178 | | 0.859 | 2.480 | 45.399 | 65 | Allowed |
| $Mg_3NAs$ | 4.236 | 4.217[35] | 0.886 | 2.119 | 36.130 | 12 | Allowed |
| $Mg_3NSb$ | 4.375 | 4.352[35] | 0.961 | 1.235 | 31.567 | 10 | Allowed |
| $Mg_3NBi$ | 4.437 | | 0.991 | 0.696 | 38.919 | 4 | Allowed |
| $Sr_3NSb$ | 5.214 | 5.173[36] | 0.915 | 0.905 | 41.695 | 14 | Allowed |
| $Ca_3NSb$ | 4.873 | 4.854[37] | 0.931 | 1.025 | 32.752 | 12 | Allowed |

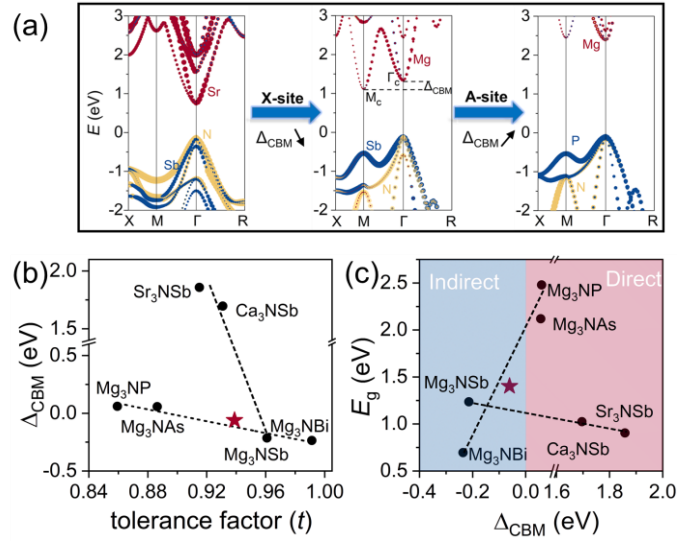

Figure 2: (a) Schematic diagram of band structure by changing the X- and A-site in antiperovskite $X_3NA$ ($X^{2+}$ = $Mg^{2+}$, $Ca^{2+}$, $Sr^{2+}$; $A^{3-}$ = $P^{3-}$, $As^{3-}$, $Sb^{3-}$, $Bi^{3-}$). Two lowest conduction band states $M_c$ and $\Gamma_c$, and the energy difference between the two states ($\Delta_{CBM}$) are labelled. (b) $\Delta_{CBM}$ as a function of tolerance factor, and (c) band gap varies as $\Delta_{CBM}$ for cubic $Pm\bar{3}m$ phase $X_3NA$ antiperovskites. The red stars in (b) and (c) show the corresponding data for $Mg_3NAs_{0.5}Bi_{0.5}$ alloy.

Taking $X_3NSb$ ($X^{2+}$ = $Mg^{2+}$, $Ca^{2+}$, $Sr^{2+}$) as an example, we analyzed the reason behind the direct-indirect transition. The variation of the two states by the orbital contribution of the two states is shown in Table 2. For $M_c$ states, the X-site contribution changes from higher energy Sr/Ca $d$ orbitals in $X_3NSb$ ($X^{2+}$ = $Sr^{2+}$ and $Ca^{2+}$) to lower Mg 3$s$ orbitals $Mg_3NSb$. As a result, the $M_c$ state is pulled down as the X-site element changes from Sr to Ca to Mg. While the higher $\Gamma_c$ state is caused by the enhanced orbital overlap between $X^{3-}$ and $Sb^{3-}$ in $Mg_3NSb$, which is consistent with the PDOS in Figure S5. This heightened orbital overlap generally correlates with the smaller X-Sb bond length. The reason for the indirect-direct transition induced by the change of the A-site composition is similar. Interestingly, this similar direct to indirect transition has also been observed in two-dimensional crystals from $Si_3O$ to $C_3O$.[38]

Table 2: The percent contributions from each atomic orbital to the two conduction band states $M_c$ and $\Gamma_c$ in $X_3NSb$ ($X^{2+}$ = $Sr^{2+}$, $Ca^{2+}$, $Mg^{2+}$).

| Compound | $\Delta_{CBM}$ (eV) | $M_c$ | | | $\Gamma_c$ | | |
|---|---|---|---|---|---|---|---|
| | | X (%) | B | N | X | B | N |
| Sr$_3$NSb | 1.858 | 50(s)+13(p)+21(d) | 10(d) | 6(s) | 13(s)+82(d) | 5(d) | 0 |
| Ca$_3$NSb | 1.696 | 44(s)+7(p)+30(d) | 11(d) | 8(s) | 11(s)+83(d) | 6(d) | 0 |
| Mg$_3$NSb | -0.214 | 63(s)+2(p) | 10(d) | 25(s) | 70(s)+13(d) | 17(d) | 0 |

To rationalize the changing trend of the $\Delta_{CBM}$ in X$_3$NA (X$^{2+}$ = Mg$^{2+}$, Ca$^{2+}$, Sr$^{2+}$; A$^{3-}$ = P$^{3-}$, As$^{3-}$, Sb$^{3-}$, Bi$^{3-}$), we summarize the $\Delta_{CBM}$ value as a function of the Goldschmidt's tolerance factor (*t*) in Figure 2b, because the *t* is an important geometric parameter to describe the structure stability of perovskite materials. Interestingly, regardless of the replacement of X-site or A-site, the $\Delta_{CBM}$ shows a negative linear dependence on the *t*, as a consequence of lower *s* orbitals of Mg and Bi elements. Therefore, we can adjust the values of $\Delta_{CBM}$ (e.g., from $\Delta_{CBM}$ < 0 eV to $\Delta_{CBM}$ > 0 eV) to optimize the corresponding indirect/direct band gap feature through choosing suitable *t*. We further establish the relationship between the band gap and the values of $\Delta_{CBM}$ in antiperovskites X$_3$NA (X$^{2+}$ = Mg$^{2+}$, Ca$^{2+}$, Sr$^{2+}$; A$^{3-}$ = P$^{3-}$, As$^{3-}$, Sb$^{3-}$, Bi$^{3-}$) in Figure 2c. It is shown that the band gap ($E_g$) decreases linearly with the $\Delta_{CBM}$ when the X site in Group II, but increases linearly with the $\Delta_{CBM}$ when changing A-site in Group VA. The positive $\Delta_{CBM}$ and optimal band gap (~ 1.5 eV) are desired for single junction solar cells. Therefore, we believe that the $\Delta_{CBM}$-$E_g$ map diagram can serve as an important guidance for designing the antiperovskite alloys in the future. Meanwhile, when changing X- or A-site elements, a linear relationship between the tolerance factor and band gaps, effective mass, and valence band width, is also observed in the studied nitride antiperovskites (see Figure S7). This is because that the *s*/*d* levels of X-site go down in energy from Mg$^{2+}$ to Sr$^{2+}$ (see Figure S8), pulling down the CBM and reducing the band gap. We note that a similar linear relationship has already been theoretically and experimentally reported in traditional inorganic halide perovskite alloys.[39, 40]

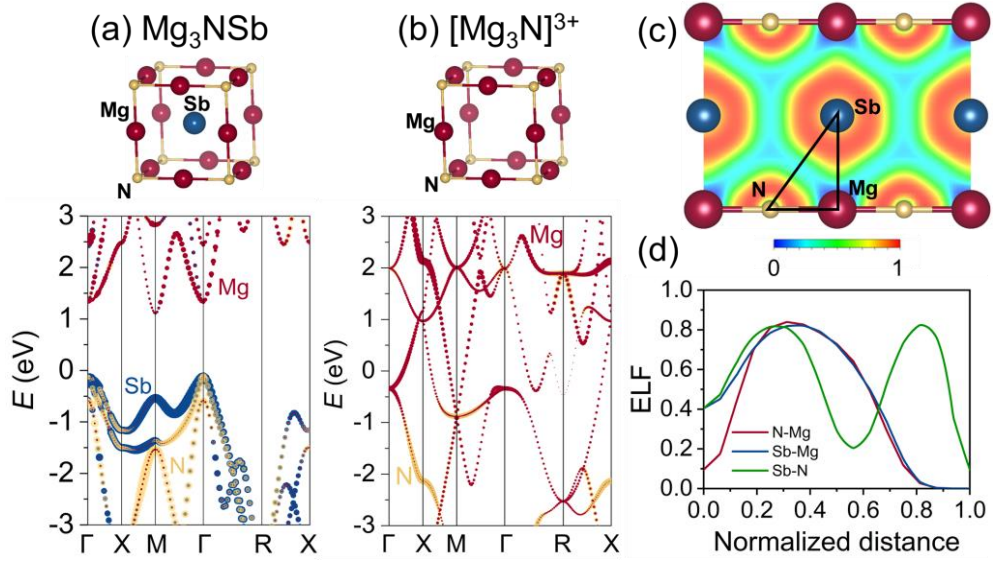

Figure 3: Comparison of band structures of antiperovskites $Mg_3NSb$ (a) and frame model of $[Mg_3N]^-$ (b) with the same cubic lattice. The VBM (Fermi level) is shifted to zero for semiconductors (metal). (c) The electron localization function (ELF) in (110) plane of cubic $Mg_3NSb$. (d) The line profile of ELF along three bonds, and the isosurface level is 0.983.

From the change of $\Delta_{CBM}$ in $Mg_3NA$ ($A^{3-}$ = $P^{3-}$, $As^{3-}$, $Sb^{3-}$, $Bi^{3-}$), it can be observed that the A-site does contribute to the electronic structure (see the PDOS in Figure S5), which is very different from the conventional halide perovskites. To further clarify the key role of A-site on the electronic structure, we construct a hypothetical frame model $[Mg_3N]^{3+}$ with $Sb^{3-}$ ions removed in the same crystal lattice (see Figure 3) and calculate its band structure. Surprisingly, the system turns out to be a metal after removing all Sb atoms, which is quite different from the original semiconducting $Mg_3NSb$, further confirming the significant contributions of A sites to the electronic structure in antiperovskites. On the contrary, in conventional halide perovskites, the band structure of $[PbI_3]^-$ model without $Cs^+$ ions is almost the same as that of the original $CsPbI_3$ (see Figure S9), indicating that the A-site cations do not affect the electronic states around the band edge. Further, from the electron localization function (ELF), it can be seen that the ELF of A-X/B-X bond in the region between the nuclei of the atoms in $Mg_3NSb$ (i.e., Sb-Ca or Sb-N, ELF > 0.2) is significantly larger than that of $CsPbI_3$ (i.e., Cs-I or Cs-Pb, ELF < 0.1 in Figure S10). This implies that the A site forms a strong bonding with other sites in the antiperovskites, and thus has a non-negligible contribution to the electronic properties accordingly.

**Static dielectric constants**

In addition to band gap, static dielectric constant is also an important descriptor for evaluating the performance of a solar cell absorber. For example, a large static dielectric constant provides strong screening, thereby suppressing carrier scattering, trapping, and recombination, eventually improving transport properties.[41] Although the electronic properties of antiperovskites have been widely reported, the dielectric properties of most of them have not yet been reported. We next focus on the static dielectric constants (ε) of $X_3NA$ ($X^{2+}$ = $Mg^{2+}$, $Ca^{2+}$, $Sr^{2+}$; $A^{3-}$ = $P^{3-}$, $As^{3-}$, $Sb^{3-}$, $Bi^{3-}$), and summarize the results in Table S3 and Figure 4. The ε values range from 31.57 in $Mg_3NSb$ to 45.40 in $Mg_3NP$, and the data of $Mg_3NSb$ is in line with previous report.[16] All the ε values are substantially enhanced compared with the calculated values in halide perovskites (i.e., $CsPbBr_3$, ~ 20.00) (see Figure 4a and Table S3),[42] which can be attributed to the larger ionic ($\varepsilon_{ion}$) and electronic ($\varepsilon_{ele}$) contributions. Here, we mainly focus on the $\varepsilon_{ele}$ that is related to the electronic properties directly. It is shown that the $\varepsilon_{ele}$ in antiperovskites $X_3NA$ is at least twice larger than those in halide perovskites. This may be due to the significant contribution of the A-site anion to the electronic properties of antiperovskites, which has been discussed before. Further, we find a positive linear relationship between $\varepsilon_{ele}$ and $t$ in $Mg_3NA$ ($A^{3-}$ = $P^{3-}$, $As^{3-}$, $Sb^{3-}$, $Bi^{3-}$), as shown in Figure 4b. With the increase of $t$, the $\varepsilon_{ele}$ increases from 9.37 at $t$ = 0.86 in $Mg_3NP$ to 20.63 at $t$ = 0.99 in $Mg_3NBi$. This is related with the linear relationship between band gap and tolerance factor in Figure S7, because a larger band gap generally leads to a smaller $\varepsilon_{ele}$.[43] On the other hand, we find that the $\varepsilon_{ele}$ is not sensitive to the X-site element because of their similar band gap values. Based on the calculated $\varepsilon_{ele}$, we estimate the exciton binding energies $E_b$ of the six studied antiperovskites $X_3NA$ (see Table 1 and Figure 4c) using the Wannier model, which is sufficient to reflect the correct trends of the the $E_b$ values (see Figure S11). $Mg_3NP$ has the largest $E_b$ of 65 meV, as a result of its largest effective mass and smallest electronic dielectric constant. The estimated $E_b$ for other five $X_3NA$ ranges from 4 meV to 14 meV (see Table 1 and Figure 4c), smaller than the thermal energy (($k_BT$ ~ 26 meV) at room temperature. These $E_b$ values are only thirty percent of those in star lead halide perovskites (i.e., CsPbBr$_3$). The small exciton binding energy in these antiperovskites will facilitate rapid electron-induced carrier dissociation, which is considerably important for an ideal solar cell absorber.

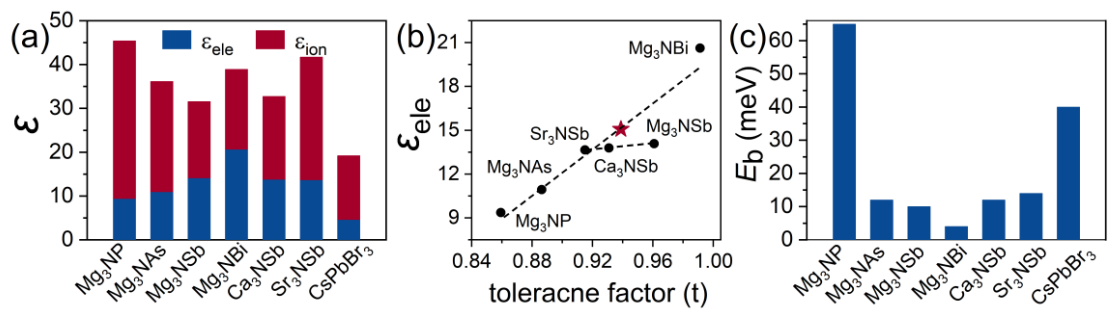

Figure 4: (a) The ionic ($\varepsilon_{ion}$) and electronic ($\varepsilon_{ele}$) contributions to the static dielectric constant ($\varepsilon_{std}$) for X$_3$NA (X$^{2+}$ = Mg$^{2+}$, Ca$^{2+}$, Sr$^{2+}$; A$^{3-}$ = P$^{3-}$, As$^{3-}$, Sb$^{3-}$, Bi$^{3-}$). (b) The electronic dielectric constant ($\varepsilon$) as a function of tolerance factor for X$_3$NA, and the red star shows the corresponding data for Mg$_3$NAs$_{0.5}$Bi$_{0.5}$ alloy. (c) The exciton binding energy ($E_b$) of X$_3$NA. The corresponding parameters in CsPbBr$_3$ are also listed for comparison.[42]

**Mechanical properties**

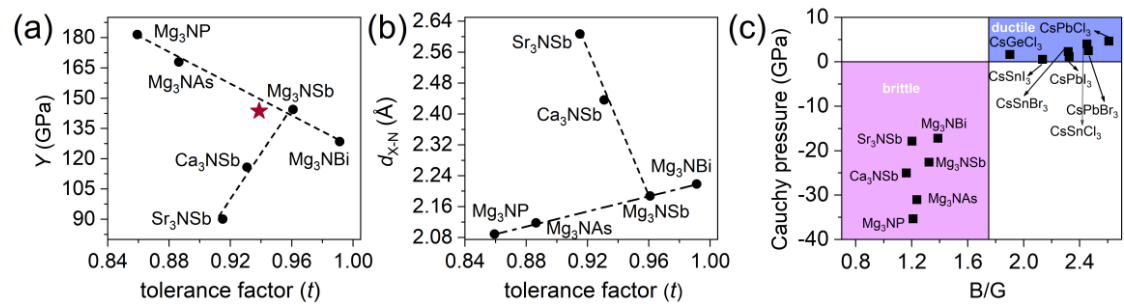

Figure 5: (a) The Young's modulus $Y$ (GPa) and (b) bond length of X-N $d_{X-N}$ (Å) as functions of tolerance factor for antiperovskites X$_3$NA (X$^{2+}$ = Mg$^{2+}$, Ca$^{2+}$, Sr$^{2+}$; A$^{3-}$ = P$^{3-}$, As$^{3-}$, Sb$^{3-}$, Bi$^{3-}$). The red star in (a) shows the corresponding data for Mg$_3$NAs$_{0.5}$Bi$_{0.5}$ alloy. (c) Correlation between Cauchy pressure and $B/G$ for all studied six nitride anti-perovskites X$_3$NA (X$^{2+}$ = Mg$^{2+}$, Ca$^{2+}$, Sr$^{2+}$; A$^{3-}$ = P$^{3-}$, As$^{3-}$, Sb$^{3-}$, Bi$^{3-}$). The data for perovskites CsBX$_3$ (B$^{2+}$ = Pb$^{2+}$, Sn$^{2+}$, Ge$^{2+}$; X$^-$ = Cl$^-$, Br$^-$, I$^-$) have been collected for comparison.

For the practical fabrication and device applications, the mechanical properties of materials are very important and they can be useful to predict their aging behaviors. We thus summarize the calculated elastic constants of all the studied antiperovskites $X_3NA$ in Table S4. The dependence of the derived Young's modulus ($Y$) on tolerance factor ($t$) is shown in Figure 5a. A similar linear relationship is clearly identified between $Y$ and $t$. For instance, for the Mg-based nitride compounds, the $Y$ value declines monotonously with increasing $t$ when changing the A-site in Group VA. The $Y$ value decreases from 181.48 GPa at $t = 0.86$ in $Mg_3NP$ to 128.49 GPa at $t = 0.99$ in $Mg_3NBi$. While for the $X_3NSb$, the $Y$ value increases monotonously with increasing $t$. When replacing the X-site in Group II, ranging from 90.10 GPa at $t = 0.915$ in $Sr_3NSb$ to 144.46 GPa at $t = 0.961$ in $Mg_3NSb$. This linear variation of Young's modulus can be related to the N-X bond length in $X_3NA$ as shown in Figure 4b, where smaller bond length generally requires large strain to break out, leading to large $Y$. Next, we elaborate the ductile or brittle nature of these antiperovskites using Cauchy pressure, which is defined as the difference between the two particular elastic constants $C_{12}$-$C_{44}$, serving as an indication of ductility. The material is expected to be brittle (ductile) for negative (positive) Cauchy pressure. Here, the Cauchy pressure for antiperovskite nitrides $X_3NA$ is negative, clearly showing the brittle nature of these compounds. The ratio between Bulk modulus $B$ and shear modulus $G$ is another index associated with the ductility nature of materials. Low (high) $B/G$ ratio suggests brittle (ductile) feature of compounds, and the critical value is 1.75.[44] This ratio ranges from 1.1 to 1.5 for the six $X_3NA$, further confirming the brittle nature of antiperovskites. The brittle nature of antiperovsites $X_3NA$ is very different from the ductile feature of common halide perovskites $CsBX_3$ ($B^{2+} = Pb^{2+}$, $Sn^{2+}$, $Ge^{2+}$; $X^- = Cl^-$, $Br^-$, $I^-$), which is clearly indicated in the map diagram in Figure 5c. Therefore, the cubic antiperovskite nitrides $X_3NA$ may possess a low probability to applicate in flexible devices in the future.

**Designing Promising Antiperovskite through A-site Alloying**

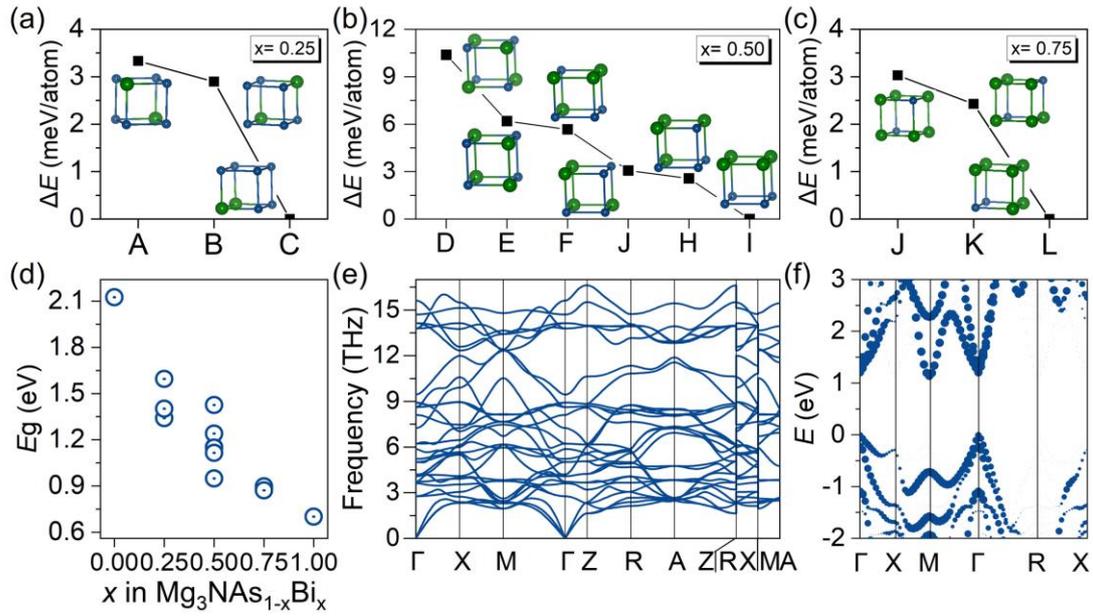

Figure 6: Energies of $Mg_3NAs_{1-x}Bi_x$ with different types of As (blue) and Bi (green) motif arrangements for (a) $x = 0.25$, (b) $x = 0.50$, and (c) $x = 0.75$. The energies of the lowest-energy configurations at each concentration are set to zero. (d) Calculated band gaps of $Mg_3NAs_{1-x}Bi_x$ as a function of $x$, the amount of Bi. (e) Phonon dispersions and (f) unfolded band structures of $Mg_3NAs_{0.5}Bi_{0.5}$ alloy in the most configuration.

The alloy strategy has been proven to be an effective way to modify the properties of perovskites and overcome some disadvantages of single compounds,[45] and the above established structure-composition-property relationship is very instructive for the design of antiperovskite alloys. Here, we investigate the electronic properties of $Mg_3NAs_{1-x}Bi_x$ alloys with $x = 0.25$, 0.50, and 0.75 for the first time, which may possess an ideal band gap (~1.5 eV) for single-junction solar cells based on the linear relationship in Figure 2c. The cubic random alloys are modeled by using 2×2×2 supercells containing 40 atoms. The Bi atoms can substitute As atoms at different positions, thus forming several different configurations with certain $x$. For example, there are 3, 6, and 3 different configurations for $x = 0.25$, 0.50, and 0.75 (see Figure 6a-c), respectively. The total energy difference between different configuration is smaller than 11 meV/atom, suggesting the possibility of forming disordered solid solutions.[46] It is shown that although the band gap differs between different configurations for a certain $x$, the band gap of $Mg_3NAs_{1-x}Bi_x$ alloys generally decreases with increasing $x$, reduced from 2.102 eV in Mg$_3$NAs to 0.612 eV in Mg$_3$NBi (see Figure 6d). This decreasing trend is associated with the higher Bi-*p* orbitals than As-*p* orbitals.

Finally, in order to validate the practicability of the structure-composition-property relationship, we focus on Mg$_3$NAs$_{0.5}$Bi$_{0.5}$ in the most stable configuration I, which has been further verified by statistical approach (see Figure S12). The configuration I shows no imaginary phonon frequencies in Figure 6e, indicating kinetic stability of the structure I. The tolerance factor of this alloy is 0.94. Figure 6f shows the band structure with the bands unfolded back from supercell to its primitive unit cell, which can be comparable with the bands of primitive cell, and quantitative analysis of alloy is enabled.[47] It clearly shows a quasi-direct band gap of 1.402 eV. This data has been added as red star sign in linear relationship in Figure S7, located at the fitted line, confirming the linear rule in band gap again. This is another confirmation that we can use this rule to design antiperovskites with optimal band gap (~1.5 eV) for solar cells. Furthermore, the $\Delta_{CBM}$ is -0.061 eV, agreeing well with the predicted data of -0.059 eV using the linear relationship in Figure 2b. For dielectric constants of Mg$_3$NAs$_{0.5}$Bi$_{0.5}$, they are isotropic in *x* and *y* directions, but a little anisotropic in *z* direction. The $\varepsilon_{ele}$ ($\varepsilon_{ion}$) is around 15.06 (23.00) in Table S3. Finally, the Young's modulus for Mg$_3$NAs$_{0.5}$Bi$_{0.5}$ is 143.61 GPa. Both the $\varepsilon_{ele}$ and *Y* are on the fitting lines in Figures 4b and 5a. Thus, the physical quantities of Mg$_3$NAs$_{0.5}$Bi$_{0.5}$ alloy confirm the uncovered structure-composition-property relationship. We can use this universal relationship to design suitable antiperovskite alloys for different optoelectronic applications.

**Comparison with lead halide perovskites**

Toward future practical application, let us now make a comprehensive comparison between antiperovskite nitrides X$_3$NA (X$^{2+}$ = Mg$^{2+}$, Ca$^{2+}$, Sr$^{2+}$; A$^{3-}$ = P$^{3-}$, As$^{3-}$, Sb$^{3-}$, Bi$^{3-}$) and the widely studied three-dimensional (3D) lead halide perovskites APbX$_3$ (A$^+$= MA$^+$, FA$^+$, Cs$^+$; X$^-$ = Cl$^-$, Br$^-$, I$^-$) (see Figure 7). Overall, X$_3$NA possess very similar electronic and optical properties as those of APbX$_3$, such as suitable band gaps (~ 1.5 eV), small carrier effect masses (0.26-0.98 m$_0$), small exciton binding energies (4-65 meV), and allowed optical transitions at band edges.

The excellent photoelectric properties can be attributed to their high symmetric crystal lattice, antibonding states of the VBM (see Figure S13), and high orbital connectivity near the band edges. Compared with APbX$_3$, the antiperovskites X$_3$NA are obtained by electronically inverting the formula, resulting in totally different band-edge characteristics. More interestingly, the effect of the A-site element on the electronic structure has a non-negligible effect in the antiperovskites, which is quite different from the halide perovskites. In addition, according to the predicted mechanical properties, antiperovskites are difficult to be applied in flexible devices in the future. On the other hand, compared with the simple low-temperature solution processing of halide perovskites, most of the antiperovskites synthesized in current experiments are powder samples, which are prepared at high temperatures, such as Mg$_3$NAs and Mg$_3$NSb.[16,35] Therefore, although the successful synthesis of the antiperovskites films in experiment has been reported, the preparation of high-quality films is still challenging, which is a key step for future device application.

Figure 7: A comprehensive comparison of 3D antiperovskite X$_3$BA and the typical 3D halide perovskite ABX$_3$ for band structure, electronic bandgap $E_g$, exciton binding energy $E_b$, carrier effective mass $m_h$, optical transition character, Young's modulus $Y$, softness, synthesis method, and sample type.

|  | 3D antiperovskite X$_3$BA | 3D halide perovskite ABX$_3$ |
|---|---|---|
| Crystal structure | 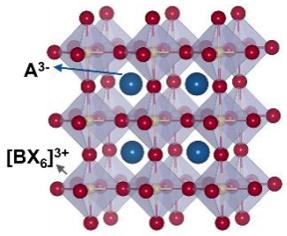 | 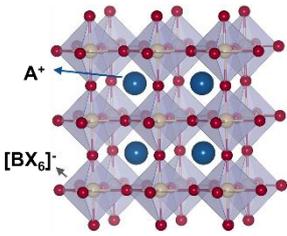 |
| Material example | A$^{3-}$ = P$^{3-}$, As$^{3-}$, Sb$^{3-}$, Bi$^{3-}$<br>B$^{3-}$ = N$^{3-}$<br>X$^{2+}$ = Mg$^{2+}$, Ca$^{2+}$, Sr$^{2+}$ | A$^+$ = MA$^+$, FA$^+$, Cs$^+$<br>B$^{2+}$ = Pb$^{2+}$<br>X$^-$ = Cl$^-$, Br$^-$, I$^-$ |
| Synthesis | Reaction tube (~800 °C) | In solution (~100 °C) |
| Processing methods | sintering, sputtering | CVD, sputtering |
| Sample types | Powder, thin film | Single crystal, thin film |

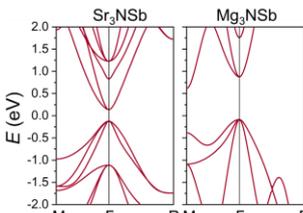

| | | |
|---|---|---|
| Typical band structures | | 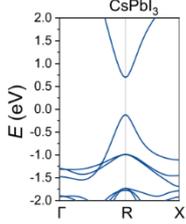 |
| Band edge's main contribution | CBM X $d/s$<br>VBM A $p$ + B $p$ | CBM B $p$<br>VBM B $s$ + X $p$ |
| $E_g$ (eV) | 0.70-2.48 | 1.40-3.21[48, 49] |
| $m_h$ ($m_0$) | 0.26-0.98 | 0.14-0.31[48, 50] |
| $E_b$ (meV) | 4-65 | 14-68[51] |
| Optical transition | Allowed | Allowed |
| $Y$ (GPa) | 90.10-181.48 | 17.79-21.92[52] |
| Brittleness/Ductility | Brittleness | Ductility |

**Conclusion**

In summary, we have studied the previously overlooked conduction band convergence, dielectric constant, and exciton binding energy of cubic $Pm\bar{3}m$ phase antiperovskite $X_3NA$ ($X^{2+}$ = $Mg^{2+}$, $Ca^{2+}$, $Sr^{2+}$; $A^{3-}$ = $P^{3-}$, $As^{3-}$, $Sb^{3-}$, $Bi^{3-}$) using first-principles calculations. Similar to lead halide perovskites, the $X_3NA$ has suitable band gaps (~ 1.5 eV), small carrier effective masses (0.26-0.98 $m_0$), small exciton binding energies (4-65 meV), and allowed optical transitions at band edges. On the other hand, $X_3NA$ exhibits totally different band edge characteristics, compared to the perovskites with electronically inverting the formula. Both the X-site and A-site can effectively tune the conduction band convergence, leading to the transition between indirect and direct band gap feature. Importantly, a universal relationship between the tolerance factor and physical quantities, including band gap, $\Delta_{CBM}$, electronic dielectric constants, and Young's modulus, is uncovered. The linear relationship originates from the atomic orbital energy of X- and A-site element. Based on the established structure-composition-property relationship in six antiperovskite nitrides $X_3NA$, we design the alloy $Mg_3NAs_{0.5}Bi_{0.5}$ with an optimal band gap of 1.402 eV as solar cell absorber. Finally, we make a comprehensive comparison between the nitride-based antiperovskites and lead halide perovskites. Our work will provide an effective strategy for designing promising antiperovskite alloys for novel device applications.

**Conflicts of interest**

There are no conflicts to declare.

**Acknowledgement**

This work is supported by the National Natural Science Foundation of China (Grant Nos. 11947218, 11704300, and 11547256), and the Natural Science Foundation of Hubei Province, China (2020CFA041). Gang Tang acknowledges the support by the Consortium des Équipements de Calcul Intensif (CÉCI) that is funded by the Fonds de la Recherche Scientifique de Belgique (F.R.S.-FNRS) under Grant No. 2.5020.11 and by the Walloon Region. Numerical calculations presented in this paper were performed on a supercomputing system in the Supercomputing Center of Wuhan University.

**Author contributions**

G.T. conceived the idea and designed the project. H.X.Z. carrier out most of the calculations and analyzed all data under S.J.Y.'s supervision. C.B.F. calculated the dielectric and mechanical properties. H.X.Z. and G.T. wrote the manuscript. S.J.Y. and G.T. revised the manuscript. All authors contributed to the discussion and revision of the paper.

**Supporting Information.** Stability, crystal and electronic structure, mechanical properties, band structure unfolding in alloy calculations, COHP analysis, structure-composition-property relationships in other antiperovskite and perovskites, and the effects of B-site elements on antiperovskites.

**References**

1. Kim, J. Y.; Lee, J.-W.; Jung, H. S.; Shin, H.; Park, N.-G., High-Efficiency Perovskite Solar Cells. *Chem. Rev.* **2020,** *120* (15), 7867-7918.
2. Tiwari, A.; Satpute, N. S.; Mehare, C. M.; Dhoble, S., Challenges, Recent Advances and Improvements for Enhancing the Efficiencies of ABX$_3$-Based PeLEDs (Perovskites Light Emitting Diodes): A Review. *J. Alloys Compd.* **2020,** *850* (0925), 156827.
3. Wang, H.; Kim, D. H., Perovskite-Based Photodetectors: Materials and Devices. *Chem. Soc. Rev.* **2017,** *46* (17), 5204-5236.
4. Fu, P.; Hu, S.; Tang, J.; Xiao, Z., Material Exploration via Designing Spatial

Arrangement of Octahedral Units: A Case Study of Lead Halide Perovskites. *Front. Optoelectron.* **2021**, 1-8.
5. Best Research-Cell Efficiency Chart. https://www:nrel:gov/pv/assets/pdfs/best-research-cell-efficiencies:20200925:pdf, Accessed October 3, 2020.
6. Kojima, A.; Teshima, K.; Shirai, Y.; Miyasaka, T., Organometal Halide Perovskites as Visible-Light Sensitizers for Photovoltaic Cells. *J. Am. Chem. Soc.* **2009,** *131* (17), 6050-6051.
7. Ju, M.-G.; Chen, M.; Zhou, Y.; Dai, J.; Ma, L.; Padture, N. P.; Zeng, X. C., Toward Eco-Friendly and Stable Perovskite Materials for Photovoltaics. *Joule* **2018,** *2* (7), 1231-1241.
8. Li, H.; Wei, Q.; Ning, Z., Toward High Efficiency Tin Perovskite Solar Cells: A Perspective. *Appl. Phys. Lett.* **2020,** *117* (6), 060502.
9. Swarnkar, A.; Mir, W. J.; Chakraborty, R.; Jagadeeswararao, M.; Sheikh, T.; Nag, A., Are Chalcogenide Perovskites an Emerging Class of Semiconductors for Optoelectronic Properties and Solar Cell? *Chem. Mater.* **2019,** *31* (3), 565-575.
10. Sun, Q.; Yin, W.-J.; Wei, S.-H., Searching for Stable Perovskite Solar Cell Materials Using Materials Genome Techniques and High-Throughput Calculations. *J. Mater. Chem. C* **2020,** *8* (35), 12012-12035.
11. Niu, S.; Sarkar, D.; Williams, K.; Zhou, Y.; Li, Y.; Bianco, E.; Huyan, H.; Cronin, S. B.; McConney, M. E.; Haiges, R., Optimal Bandgap in A 2D Ruddlesden–Popper Perovskite Chalcogenide for Single-Junction Solar Cells. *Chem. Mater.* **2018,** *30* (15), 4882-4886.
12. Wang, Y.; Zhang, H.; Zhu, J.; Lü, X.; Li, S.; Zou, R.; Zhao, Y., Antiperovskites with Exceptional Functionalities. *Adv. Mater.* **2020,** *32* (7), 1905007.
13. Gebhardt, J.; Rappe, A. M., Adding to the Perovskite Universe: Inverse-Hybrid Perovskites. *ACS Energy Lett.* **2017,** *2* (12), 2681-2685.
14. Gebhardt, J.; Rappe, A. M., Design of Metal-Halide Inverse-Hybrid Perovskites. *J. Phys. Chem. C* **2018,** *122* (25), 13872-13883.
15. Wang, Z.-X.; Zhang, Y.; Tang, Y.-Y.; Li, P.-F.; Xiong, R.-G., Fluoridation Achieved Antiperovskite Molecular Ferroelectric in $[(CH_3)_2(F-CH_2CH_2)NH]_3(CdCl_3)(CdCl_4)$. *J. AM. Chem. Soc.* **2019,** *141* (10), 4372-4378.
16. Heinselman, K. N.; Lany, S.; Perkins, J. D.; Talley, K. R.; Zakutayev, A., Thin Film Synthesis of Semiconductors in the Mg–Sb–N Materials System. *Chem. Mater.* **2019,** *31* (21), 8717-8724.
17. Dai, J.; Ju, M.-G.; Ma, L.; Zeng, X. C., Bi(Sb)NCa$_3$: Expansion of Perovskite Photovoltaics into All-Inorganic Anti-Perovskite Materials. *J. Phys. Chem. C* **2019,** *123* (11), 6363-6369.
18. Mochizuki, Y.; Sung, H.-J.; Takahashi, A.; Kumagai, Y.; Oba, F., Theoretical Exploration of Mixed-Anion Antiperovskite Semiconductors M$_3$XN (M= Mg, Ca, Sr, Ba; X= P, As, Sb, Bi). *Phys. Rev. Mater.* **2020,** *4* (4), 044601.
19. Kresse, G.; Furthmüller, J., Efficiency of Ab-Initio Total Energy Calculations for Metals and Semiconductors Using A Plane-Wave Basis Set. *Comput. Mater. Sci.* **1996,** *6* (1), 15-50.
20. Kresse, G.; Joubert, D., From Ultrasoft Pseudopotentials to the Projector Augmented-Wave Method. *Phys. Rev. B* **1999,** *59* (3), 1758.
21. Blöchl, P. E., Projector Augmented-Wave Method. *Phys. Rev. B* **1994,** *50* (24), 17953.

22. Perdew, J. P.; Burke, K.; Ernzerhof, M., Generalized Gradient Approximation Made Simple. *Phys. Rev. Lett.* **1996,** *77* (18), 3865.

23. Heyd, J.; Peralta, J. E.; Scuseria, G. E.; Martin, R. L., Energy Band Gaps and Lattice Parameters Evaluated with the Heyd-Scuseria-Ernzerhof Screened Hybrid Functional. *J. Chem. Phys.* **2005,** *123* (17), 174101.

24. Monkhorst, H. J.; Pack, J. D., Special Points for Brillouin-Zone Integrations. *Phys. Rev. B* **1976,** *13* (12), 5188.

25. Togo, A.; Oba, F.; Tanaka, I., First-Principles Calculations of the Ferroelastic Transition between Rutile-Type and CaCl$_2$-Type SiO$_2$ at High Pressures. *Phys. Rev. B* **2008,** *78* (13), 134106.

26. Wang, V.; Xu, N.; Liu, J.-C.; Tang, G.; Geng, W.-T., VASPKIT: A User-Friendly Interface Facilitating High-Throughput Computing and Analysis Using VASP Code. *Comput. Phys. Commun.* **2021**, 108033.

27. Ravindran, P.; Fast, L.; Korzhavyi, P. A.; Johansson, B.; Wills, J.; Eriksson, O., Density Functional Theory for Calculation of Elastic Properties of Orthorhombic Crystals: Application to TiSi$_2$. *J. Appl. Phys.* **1998,** *84* (9), 4891-4904.

28. Baroni, S.; De Gironcoli, S.; Dal Corso, A.; Giannozzi, P., Phonons and Related Crystal Properties from Density-Functional Perturbation Theory. *Rev. Mod. Phys.* **2001,** *73* (2), 515.

29. Gonze, X.; Lee, C., Dynamical Matrices, Born Effective Charges, Dielectric Permittivity Tensors, and Interatomic Force Constants from Density-Functional Perturbation Theory. *Phys. Rev. B* **1997,** *55* (16), 10355.

30. Ghasemi, S.; Alihosseini, M.; Peymanirad, F.; Jalali, H.; Ketabi, S.; Khoeini, F.; Neek-Amal, M., Electronic, Dielectric, and Optical Properties of Two-Dimensional and Bulk Ice: A Multiscale Simulation Study. *Phys. Rev. B* **2020,** *101* (18), 184202.

31. Zhao, X.-G.; Yang, J.-H.; Fu, Y.; Yang, D.; Xu, Q.; Yu, L.; Wei, S.-H.; Zhang, L., Design of Lead-Free Inorganic Halide Perovskites for Solar Cells via Cation-Transmutation. *J. Am. Chem. Soc.* **2017,** *139* (7), 2630-2638.

32. Berger, R. F., Design Principles for the Atomic and Electronic Structure of Halide Perovskite Photovoltaic Materials: Insights from Computation. *Chem. Eur. J.* **2018,** *24* (35), 8708-8716.

33. Tang, G.; Ghosez, P.; Hong, J., Band-Edge Orbital Engineering of Perovskite Semiconductors for Optoelectronic Applications. *J. Phys. Chem. Lett.* **2021,** *12*, 4227-4239.

34. Tang, G.; Xiao, Z.; Hong, J., Designing Two-Dimensional Properties in Three-Dimensional Halide Perovskites via Orbital Engineering. *J. Phys. Chem. Lett.* **2019,** *10* (21), 6688-6694.

35. Chi, E.; Kim, W.; Hur, N.; Jung, D., New Mg-Based Antiperovskites PnNMg$_3$ (Pn= As, Sb). *Solid State Commun.* **2002,** *121* (6-7), 309-312.

36. Gäbler, F.; Kirchner, M.; Schnelle, W.; Schwarz, U.; Schmitt, M.; Rosner, H.; Niewa, R., (Sr$_3$N)E and (Ba$_3$N)E (E= Sb, Bi): Synthesis, Crystal Structures, and Physical Properties. *Z. Anorg. Allg. Chem.* **2004,** *630* (13-14), 2292-2298.

37. Chern, M. Y.; Vennos, D.; DiSalvo, F., Synthesis, Structure, and Properties of Anti-Perovskite Nitrides Ca$_3$MN, M= P, As, Sb, Bi, Ge, Sn, and Pb. *J. Solid State Chem.* **1992,** *96* (2), 415-425.

38. Chae, K.; Son, Y.-W., A New Family of Two-Dimensional Crystals: Open-Framework $T_3X$ (T= C, Si, Ge, Sn; X= O, S, Se, Te) Compounds with Tetrahedral Bonding. *Nano Lett.* **2019,** *19* (4), 2694-2699.

39. Linaburg, M. R.; McClure, E. T.; Majher, J. D.; Woodward, P. M., $Cs_{1-x}Rb_xPbCl_3$ and $Cs_{1-x}Rb_xPbBr_3$ Solid Solutions: Understanding Octahedral Tilting in Lead Halide Perovskites. *Chem. Mater.* **2017,** *29* (8), 3507-3514.

40. Lee, J.-H.; Bristowe, N. C.; Lee, J. H.; Lee, S.-H.; Bristowe, P. D.; Cheetham, A. K.; Jang, H. M., Resolving the Physical Origin of Octahedral Tilting in Halide Perovskites. *Chem. Mater.* **2016,** *28* (12), 4259-4266.

41. Ming, W.; Shi, H.; Du, M.-H., Large Dielectric Constant, High Acceptor Density, and Deep Electron Traps in Perovskite Solar Cell Material $CsGeI_3$. *J. Mater. Chem. A* **2016,** *4* (36), 13852-13858.

42. Han, D.; Shi, H.; Ming, W.; Zhou, C.; Ma, B.; Saparov, B.; Ma, Y.-Z.; Chen, S.; Du, M.-H., Unraveling Luminescence Mechanisms in Zero-Dimensional Halide Perovskites. *J. Mater. Chem. C* **2018,** *6* (24), 6398-6405.

43. Takahashi, A.; Kumagai, Y.; Miyamoto, J.; Mochizuki, Y.; Oba, F., Machine Learning Models for Predicting the Dielectric Constants of Oxides Based on High-Throughput First-Principles Calculations. *Phys. Rev. Mater.* **2020,** *4* (10), 103801.

44. Elahmar, M.; Rached, H.; Rached, D.; Khenata, R.; Murtaza, G.; Omran, S. B.; Ahmed, W., Structural, Mechanical, Electronic and Magnetic Properties of a New Series of Quaternary Heusler Alloys CoFeMnZ (Z= Si, As, Sb): a First-Principle Study. *J. Magn. Magn. Mater.* **2015,** *393*, 165-174.

45. Dalpian, G. M.; Zhao, X.-G.; Kazmerski, L.; Zunger, A., Formation and Composition-Dependent Properties of Alloys of Cubic Halide Perovskites. *Chem. Mater.* **2019,** *31* (7), 2497-2506.

46. Han, D.; Feng, C.; Du, M.-H.; Zhang, T.; Wang, S.; Tang, G.; Bein, T.; Ebert, H., Design of High-Performance Lead-Free Quaternary Antiperovskites for Photovoltaics via Ion Type Inversion and Anion Ordering. *J. AM. Chem. Soc.* **2021,** *143* (31), 12369-12379.

47. Tan, Y.; Chen, F. W.; Ghosh, A. W., First Principles Study and Empirical Parametrization of Twisted Bilayer $MoS_2$ Based on Band-Unfolding. *Appl. Phys. Lett.* **2016,** *109* (10), 101601.

48. Blancon, J.-C.; Even, J.; Stoumpos, C. C.; Kanatzidis, M. G.; Mohite, A. D., Semiconductor Physics of Organic–Inorganic 2D Halide Perovskites. *Nat. Nanotechnol.* **2020,** *15* (12), 969-985.

49. Ravi, V. K.; Markad, G. B.; Nag, A., Band Edge Energies and Excitonic Transition Probabilities of Colloidal $CsPbX_3$ (X= Cl, Br, I) Perovskite Nanocrystals. *ACS Energy Lett.* **2016,** *1* (4), 665-671.

50. Wang, S.; Xiao, W.-b.; Wang, F., Structural, Electronic, and Optical Properties of Cubic Formamidinium Lead Iodide Perovskite: A First-Principles Investigation. *RSC Adv.* **2020,** *10* (54), 32364-32369.

51. Yang, Z.; Surrente, A.; Galkowski, K.; Bruyant, N.; Maude, D. K.; Haghighirad, A. A.; Snaith, H. J.; Plochocka, P.; Nicholas, R. J., Unraveling the Exciton Binding Energy and the Dielectric Constant in Single-Crystal Methylammonium Lead Triiodide Perovskite. *J. Phys. Chem. Lett.* **2017,** *8* (8), 1851-1855.

52. Faghihnasiri, M.; Izadifard, M.; Ghazi, M. E., DFT Study of Mechanical Properties and

Stability of Cubic Methylammonium Lead Halide Perovskites ($CH_3NH_3PbX_3$, X= I, Br, Cl). *J. Phys. Chem. C* **2017,** *121* (48), 27059-27070.